\documentclass[conference]{IEEEtran}
\IEEEoverridecommandlockouts
\usepackage[top=0.76in, bottom=1.1in, left=0.625in, right=0.625in]{geometry}
\usepackage{cite}
\usepackage{amsmath,amssymb,amsfonts}
\usepackage{graphicx}
\usepackage{textcomp}
\usepackage{xcolor}
\usepackage{physics}
\usepackage{asymptote}
\usepackage{standalone}
\usepackage{tikz}
\usepackage[inkscapelatex=false]{svg}
\usepackage{fix-cm}
\usepackage[caption=false,font=footnotesize]{subfig}
\usetikzlibrary{arrows.meta, positioning, fit, backgrounds}
\usepackage{algpseudocode}
\usepackage[linesnumbered,ruled,vlined]{algorithm2e}
\SetKwInput{KwInput}{Inputs}                
\SetKwInput{KwOutput}{Outputs}
\SetKwInput{KwInit}{Initialization}

\def\BibTeX{{\rm B\kern-.05em{\sc i\kern-.025em b}\kern-.08em
    T\kern-.1667em\lower.7ex\hbox{E}\kern-.125emX}}
\begin{document}

\title{Near-field Wideband Multi-User Localization using NFMR-Net}

\author{Pearl Hetul Shah, Srikar Sharma Sadhu, and  Praful D. Mankar
\thanks{P. H. Shah,  S. S. Sadhu,  and P. D. Mankar are with 
Signal Processing and Communication Research Center, IIIT Hyderabad, India. 
Email: pearl.shah@students.iiit.ac.in,  srikar.sadhu@research.iiit.ac.in 
and praful.mankar@iiit.ac.in. 
} 
}

\maketitle

\begin{abstract}
This paper proposes a deep learning based method for wideband near-field multi-user localization. In particular, the proposed approach utilizes the Zadoff-Chu (ZC) sequence based pilots to mitigate the inter-user interference, which in turn aids the estimation of the multi-tap channel matrix. From this channel matrix, we extract the line-of-sight (LoS) array response based on the delay-tap energy profile. The LoS delay-tap is further refined using parabolic interpolation to obtain the coarse estimate of range parameter. Next, the extracted LoS array response is used to obtain the coarse angle estimate using 2D MUSIC algorithm. These coarse estimates are further refined using the near-field music refinement network (NFMR-Net), which involves separate sub-networks for range and angle estimations. Through numerical analysis, the proposed NFMR-Net is demonstrated to outperform conventional 2D MUSIC algorithm.
\end{abstract}


\vspace{-3mm}
\section{Introduction}
The sixth generation (6G) network is expected to enable a wide range of location aware applications, including augmented/virtual reality, autonomous driving and delivery, industry automatization, etc., \cite{Zhao_6G_white_paper_near_field}. These  applications often demand high precision localization capabilities to function effectively.

The user localization is traditionally  considered for the far-field scenario, wherein the channel response vector is  characterized by the angle of arrival.  However, it is expected that the 6G  network will offer more denser access points that are equipped with  large antenna arrays operating at high frequency, which in turn will lead to  ubiquitous near-field scenario~\cite{Liu_2023}. 
While there have been significant  efforts recently in near-field localization, the conventional approaches are often designed under certain assumptions and their ability to perform precise positioning is also limited due to the nature of near-field channel. This is particularly inevitable in the presence of multipath components and multi-user interference.
Leveraging their ability to handle system level non-linearities, machine learning (ML) models can be utilized to  learn the local signatures of multipath channel behavior in the near-field and aid to perform robust localization.   
Hence, this paper focuses on exploring ML models for near-field localization.

\emph{Related Works:}
Localization has been an active area of research for several decades, continually improving  the precision with modernization of the wireless network. 
Early efforts focused on designing localization methods for far-field scenario utilizing the received signal strengths (RSS), time of arrivals (ToA), and angle of arrivals (AoA) information acquired using multi-antenna systems and/or multiple cooperative anchor nodes. The interested readers may refer to comprehensive overview and survey given in articles \cite{8409950,7762095}. 
However, as mentioned above, the near-field localization is becoming crucial for enabling  6G  applications.

The authors of \cite{Wang_subspace,luo2023beam} present near-field  localization methods in the absence of multipath fading.
The authors of \cite{Wang_subspace} apply MUSIC for multi-user localization for narrowband system. In particular, they leveraged the symmetry of the uniform linear array (ULA) to decouple the 2D MUSIC problem into separate 1D searches for range and angle estimations.
The authors of \cite{luo2023beam} exploit the beam squint phenomenon of a wideband massive multiple-input-multiple-output (MIMO) system to sequentially estimate the location of the multiple-users in the near-field.  

These conventional approaches usually rely on assumptions and approximations,  such as no multipath fading or  Fresnel approximation to make the channel response vector tractable, which may affect the overall precision. 
For these reasons,  the ML methods are  utilized in literature to handle the non-linearities in the near-field channel response.  
For example, \cite{gast2025nearfieldlocalizationaiaided} presents an ML-based approach for near-field localization of multiple users with correlated signals. The authors first  obtain a surrogate covariance matrix using deep leaning model that has a distinctly identifiable eigenspectrum. Next, 2D MUSIC is applied on this surrogate covariance matrix to estimate the locations.  
Further, a convolutional neural network (CNN) based wideband localization method is presented in  \cite{NF_AI_1} to address the beam squint effects and spatial non-stationarity of the near-field channels. The authors  first determine the coarse estimates of angle and range using controllable beam squint-based beam training. These estimates are further refined using  a CNN model that leverages large kernels to mitigate the effects of spatial non-stationarity. 
The above works consider the line-of-sight (LoS) channel model while ignoring the non-LoS (NLoS) components. This limits these works from the deployment in  practical scenarios, which involves unavoidable NLoS paths.

The authors of \cite{7102989} present the relevance vector machine based  techniques  to identify and mitigate the NLoS components for obtaining the accurate and robust ToA estimates for ultra-wideband localization.
The authors of \cite{Lu_DNOMP} present the near-field localization using extremely large antenna array  system in the presences of multipath scattering. Exploiting the spatial domain sparsity, the authors present a damped Newtonized orthogonal matching pursuit  algorithm that first apply dictionary learning to detect a new path and obtain coarse estimates of its range, azimuth and elevation  AoAs, and finally  refines these estimates using damped Newton refinement method. 
Further, a two-stage beamspace MUSIC method is proposed in \cite{10547324} for near-field localization using XL-MIMO systems. An optimally pre-compensated distance based codebook is applied  enable  1D sequential searches for angle  and  distance parameter estimations. The authors employed spatial smoothing to mitigate effect of multipath coherence.
While the above works consider NLoS paths, they are limited to the single user localization in the near-field. 
However, to the best of authors knowledge, the ML approach has  not been explored for multi-user localization in the near-field scenario while accounting for the  multipath scattering, which is the objective of this paper.

\emph{Contributions:}
This paper propose a deep-learning framework for wideband multi-user localization in the near-field. The contributions of this paper are summarized below.  
\begin{enumerate}
    \item Exploiting the low autocorrelation of Zadoff-Chu (ZC) pilots, we estimate each user's multi-tap channel matrix which allows the  extraction of LoS array response and its corresponding delay tap. 
    \item LoS response and its associated delay are utilized to obtain coarse estimates of range and angle using parabolic interpolation and 2D MUSIC.
    \item NFMR-Net is proposed for correction of residual error in coarse estimates. NFMR-Net utilizes feature vectors that captures local range and angle signatures obtained from LoS response vector and MUSIC spectrum. 
    \item Numerical results demonstrate that the proposed NFMR-Net provides a significantly lower  root mean square error (RMSE) compared to  2D MUSIC algorithm.
\end{enumerate}

\emph{Notation:}
Vectors are denoted using bold lowercase letters (e.g. $\mathbf{q}$) and matrices are denoted using bold uppercase letters (e.g. $\mathbf{Q}$) with $(\cdot)^H$ as their conjugate transpose. 
The $i$-th element of a vector $\mathbf{q}$ and $(i,j)$-th element of a matrix $\mathbf{Q}$ are $\mathbf{q}[i]$ and $\mathbf{Q}[i,j]$ respectively. 
\vspace{-1mm}
\section{System Model}
\label{sec:system_model}
\begin{figure}[t!]
    \centering
    \vspace{2mm}
    \includegraphics[width=0.75\linewidth]{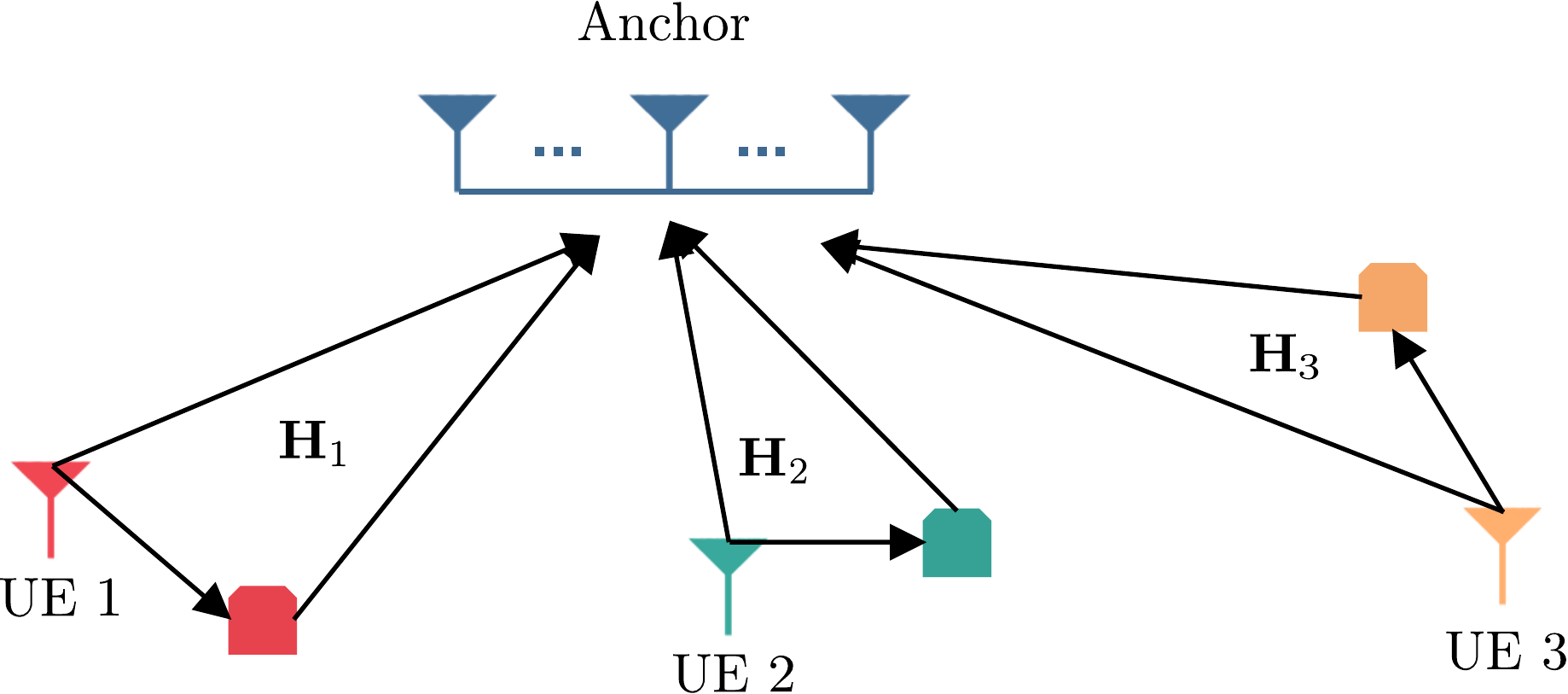}
    \caption{An Illustration of the System Model}
    \label{fig:system_model}
    \vspace{-4mm}
\end{figure}

We consider a wideband uplink positioning system in which $K$ single-antenna user equipment (UE) nodes at unknown positions transmit
simultaneously to an anchor node placed at the origin. The anchor node is assumed to be equipped with ULA placed along $x$-axis, as shown in   Fig. \ref{fig:system_model}. The UEs are considered to be in the near-field of anchor's ULA that consists of $M$ antennas with inter-element spacing
$d_A = \lambda/2$, where $\lambda$ is the carrier wavelength. The $m$-th element on the ULA is assumed to be located at $\mathbf{p}_m = [(m - (M-1)/2)\,d_A, 0]^T$, for $m = 0, 1, \dots, M-1$. 

The  location of $k$-th UE ($k = 1,\dots,K$) is denoted by $\mathbf{u}_k$.
The wideband multipath channel from the $k$-th UE to the anchor consists of both a LoS path and $P_k$ NLoS
paths, where the $p$-th scatterer of the
$k$-th UE is located at $\mathbf{s}_{k,p}$. It is to be noted that the scattering locations and number of scatterers are not considered to be known.
The positions of the $k$-th UE and its $p$-th scatterer are denoted respectively using
$\mathbf{u}_k = \left[r_k\cos\theta_k,\, r_k\sin\theta_k\right]^T$ and $\mathbf{s}_{k,p} = \left[r_{k,p}\cos\theta_{k,p},\, r_{k,p}\sin\theta_{k,p}\right]^T$.

The channel is modeled over $L$ discrete delay taps with sampling period
$T_s = 1/B$, where $B$ is the system bandwidth. Here, $L$ depends on the near-field boundary, i.e. Rayleigh distance $R_{\rm Ray}=\frac{2(Md_A)^2}{\lambda}$, such that $L=\frac{1.5R_{\rm Ray}}{c T_s}$. 
The path length resolution is $cT_s$, meaning the path with length $r$ will arrive at receiver with corresponding delay tap $\lceil\frac{r}{cT_s}\rceil$.
Thus, $k$-th UE's channel matrix
$\mathbf{H}_k \in \mathbb{C}^{M \times L}$ maps these propagation paths
onto their respective delay taps according to the multipath distances as
\begin{equation}\label{eq:Hk}
    \mathbf{H}_k = \mathbf{H}_{k,0} + \sum \nolimits_{p=1}^{P_k} \mathbf{H}_{k,p},
\end{equation}
where $\mathbf{H}_{k,0}$ and $\mathbf{H}_{k,p}$  are LoS and $p$-th NLoS channel matrices, respectively. These matrices can be modeled as
\begin{align*}
    \mathbf{H}_{k,0} &= \tfrac{1}{r_k}\mathbf{a}_{\rm L}(r_k,\theta_k)\delta\!\left(\ell - \ell_{k,0}\right), \text{~and}\\
    \mathbf{H}_{k,p} &= \tfrac{\beta}{r_{k,p}}\,
        \phi_{k,p}\mathbf{a}_{\rm NL}(r_{k,p},\theta_{k,p})
        \delta\!\left(\ell - \ell_{k,p}\right),
\end{align*}

where $\phi_{k,p}=\exp(-j\kappa\norm{\mathbf{s}_{k,p} - \mathbf{u}_{k}})$, $\kappa=\frac{2\pi}{\lambda}$, and  $\beta$  is the absorption loss parameter, respectively, from $k$-th UE to anchor. Here, $\delta(\ell - \ell') =1$ if $\ell = \ell'$, zero otherwise for $\ell,\ell'=0,\dots,L-1$. The LoS and NLoS delay tap indices are denoted as $\ell_{k,0} = \lceil r_k/(c\,T_s)\rceil$ and $\ell_{k,p} = \lceil r_{k,p}/(c\,T_s)\rceil$ respectively.
Further, $\mathbf{a}_{\rm L}(\cdot,\cdot)$ and $\mathbf{a}_{\rm NL}(\cdot,\cdot)$ denote LoS and NLoS near-field beam steering vectors respectively such that 
\begin{align*}
    \mathbf{a}_{\rm L}(r_k,\theta_k)[m] &= \exp\left(-j\kappa\norm{\mathbf{p}_m - \mathbf{u}_k}\right),\text{~~and}\\
    \mathbf{a}_{\rm NL}(r_{k,p},\theta_{k,p})[m] &= \exp\left(-j\kappa\norm{\mathbf{p}_m - \mathbf{s}_{k,p}}\right).
\end{align*} 

To enable extraction of   channels $\mathbf{H}_k$s, each UE is assigned with ZC  pilot sequence of length $N$ having a distinct root $z_k$ \cite{andrews2025primerzadoffchusequences}. The ZC pilot sequence  assigned to $k$-th UE is
\begin{equation}
\label{eq:zc}
    \mathbf{x}_k[n] = e^{-j\pi z_k \frac{n^2}{N}},
    \quad n = 0, 1, \dots, N - 1.
\end{equation}
The roots $\{z_k\}_{k=1}^{K}$ are chosen to be mutually
coprime integers. The $k$-th UE is assumed to transmit the sequence $x_k$ after adding cyclic prefix (CP) of length at least $L-1$. 
Thus, after CP removal at the receiver, we have
\begin{equation}
\label{eq:circ_conv}
\begin{aligned}
\mathbf{Y}_k[m,n] &= \sum \nolimits_{\ell=0}^{L-1} \mathbf{H}_k[m,\ell]\;
\mathbf{x}_k\!\left[[n-\ell]_{N}\right] + \mathbf{w}[m,n],
\end{aligned}
\end{equation}
for $n=0,\dots,N-1$ where $[n-\ell]_{N}= (n-\ell)\bmod N$.
Stacking \eqref{eq:circ_conv} across all time samples $n$ and all
antenna indices $m$, the contribution of $k$-th UE can be expressed in
matrix form as $\mathbf{H}_k \mathbf{X}_k$, where
$\mathbf{H}_k \in \mathbb{C}^{M\times L}$ contains the channel taps
and $\mathbf{X}_k \in \mathbb{C}^{L\times N}$ is the circulant pilot
matrix defined by
{\small\begin{equation*}
    \mathbf{X}_k=\begin{bmatrix}
    x_k[0]       & x_k[1] & \dots & x_k[N-1] \\
    x_k[N-1]   & x_k[0] & \dots & x_k[N-2] \\
    \vdots       & \vdots & \dots & \vdots\\
    x_k[N-L+1] & x_k[N-L+2] & \dots &  x_k[N-L]
\end{bmatrix}.
\end{equation*}}
It may be noted that the $(\ell+1)$-th row of $\mathbf{X}_k$ corresponds to a circular shift of sequence $x_k$ by $\ell$ samples.  We assume that each UE transmits the pilot sequence $\mathbf{X}_k$ for $T$ times over the same channel. The  received signal under $t$-th transmission is
\begin{equation}
\label{eq:Y}
\mathbf{Y}^{t}
=
\sum \nolimits_{k=1}^{K}\mathbf{H}_k\mathbf{X}_k+\mathbf{W}^{t},
\end{equation}
where $\mathbf{W}^{t}\in\mathbb{C}^{M\times N}$ denotes i.i.d.\
additive white Gaussian noise with variance $\sigma_n^2$.
Using the cyclic correlation properties of ZC sequences \cite{andrews2025primerzadoffchusequences}, the normalized
correlation matrix between two circulant pilot matrices can be written as
{\small\begin{equation}
\label{eq:corr_cases}
\frac{1}{N}\mathbf{X}_j\mathbf{X}_k^H
=
\begin{cases}
\mathbf{I}_L, & j = k, \\[1mm]
\frac{1}{\sqrt{N}}\mathbf{\Psi}_{jk}, & j \neq k,
\end{cases}
\end{equation}}
where $\mathbf{I}_L$ is $L\times L$ identity matrix and  $\mathbf{\Psi}_{jk}\in\mathbb{C}^{L\times L}$ satisfying
$|[\mathbf{\Psi}_{jk}]_{\ell,\ell'}| = 1$ for $\forall\,(\ell,\ell')$. 
For $j=k$, each row of $\mathbf{X}_k$ is a circular shift of the same
ZC sequence and thus the inner product of its rows follow its cyclic autocorrelation. Hence, the correlation equals $N$ for zero shift (i.e. same row indices) and zero for any non-zero shift (i.e. different row indices). This
yields $\frac{1}{N}\mathbf{X}_k\mathbf{X}_k^H=\mathbf{I}_L$. For $j\neq k$, pilots are generated from distinct coprime roots, and
their cyclic cross-correlation is constant in magnitude. 

\section{Proposed Multi-Stage Estimation Framework}
\label{sec:proposed_method}

The proposed framework estimates the location  of
each UE $(r_k, \theta_k)$ in four sequential stages: (i)~extraction of  array response corresponding to LoS tap 
(ii)~sub-tap range refinement via parabolic interpolation
(iii)~evaluate 
2D-MUSIC spectrum over the near-field local grid
(iv)~joint $(r,\theta)$ refinement via \emph{NFMR-Net}.
These steps are discussed in detail in this section.

\subsection{Extraction of LoS Array Response}
\label{subsec:channel_estimation}
We first estimate the channel matrix corresponding to $k$-th UE by correlation property of ZC sequences as
\begin{align}
\hat{\mathbf{H}}_k^{t}&=\frac{1}{N}\mathbf{Y}^{t}\mathbf{X}_k^H
\stackrel{(a)}{=}\mathbf{H}_k+\frac{1}{\sqrt{N}} \sum_{j\neq k}\mathbf{H}_j\mathbf{\Psi}_{jk}+\tilde{\mathbf{W}^{t}},\label{eq:Hk_hat}
\end{align}
where $\tilde{\mathbf{W}^{t}}=\frac{1}{N}\mathbf{W}^{t}\mathbf{X}_k^H$ and the step (a) follows using \eqref{eq:corr_cases}.
In the above equation,  second term represents the multi-user interference (MUI) whose magnitude is bounded by $1/\sqrt{N}$ and the third term corresponds to post-correlation noise  with variance $\tfrac{\sigma_n^2}{N}$. Thus, for sufficiently large values of $N$ and SNR, \eqref{eq:Hk_hat} provides an accurate estimate of $k$-th UE's channel matrix.

We assume that the LoS response is dominant compared to the NLoS multipath components, i.e. $|\beta_{k,p}|^2 \ll |\beta_{k,0}|^2$. Thus, we use the tap-wise energies to identify the  delay tap corresponding to the LoS component.  The tap energy profile is averaged over $T$ snapshots to obtain a cleaner estimate:
\begin{equation}
\label{eq:tap_energy}
e_k[\ell]
=
\tfrac{1}{TM}
{\sum \nolimits_{t=1}^{T}}\|\hat{\mathbf{H}}_k^{(t)}[:,\ell]\|^2,
\quad \ell=0,\dots,L-1.
\end{equation}
The LoS delay tap of the $k$-th UE is estimated as
\begin{equation}
\label{eq:los_tap}
\hat{\ell}_k
=
\operatorname*{arg\,max}_{\ell\in\mathcal{L}_k}
e_k[\ell],
\end{equation}
where $\mathcal{L}_k$ contains indices satisfying the local peak condition $e_k[\ell]\ge e_k[\ell\pm1]$ and the threshold condition
$e_k[\ell]\ge\epsilon \max_{\ell'} e_k[\ell']$. This search space is constrained by the parabolic interpolation which we use to estimate the range, as will be discussed shortly. 
 Using $\hat{\ell}_k$, the LoS array response vector for $k$-th UE can be extracted as $\hat{\mathbf{h}}_k^{t} = \hat{\mathbf{H}}_k^{t}[:,\hat{\ell}_k]$  for snapshot $t$. Let $\tilde{\mathbf{w}}^t=\tilde{\mathbf{W}}^t[:,\hat{\ell}_k]$. Thus, we have
\begin{equation}
\hat{\mathbf{h}}_k^t=\beta_{k,0}{\mathbf{a}}_{\rm L}(r_k,\theta_k)+\tilde{\mathbf{w}}^t. \label{eq:h_k_hat}   
\end{equation}
\subsection{Sub-Tap Range Refinement}
\label{subsec:parabolic}
Using the estimate of LoS delay tap, we can determine the  estimate of the range parameter of the $k$-th UE as $$\hat{r}_k^{c} =
\hat{\ell}_k \cdot c\,T_s.$$ 
However, this estimate involves a quantization error that is bounded by
$c\,T_s/2$, which is half of the distance resolution interval. To
suppress this error without increasing the bandwidth (or, equivalently reducing the sample interval), we apply
parabolic interpolation~\cite{9325013} on the three-point neighborhood of the coarse estimate of the delay tap $\hat{\ell}_k$.
Assuming the tap energy profile is locally smooth near the true delay, the detected peak is approximated by a quadratic curve
fitted through the three adjacent samples
$\{e_k[\hat{\ell}_k-1],\; e_k[\hat{\ell}_k],\; e_k[\hat{\ell}_k+1]\},$
and the vertex location (i.e., true LoS delay) provides a fractional correction to the tap index (i.e., estimated delay tap $\hat{\ell}_k$). The correction is
\begin{equation}
\label{eq:parabolic}
    \delta_k = \frac{1}{2}\cdot
    \frac{e_k[\hat{\ell}_k - 1] - e_k[\hat{\ell}_k + 1]}
         {e_k[\hat{\ell}_k - 1] - 2\,e_k[\hat{\ell}_k]
          + e_k[\hat{\ell}_k + 1]},
\end{equation}
with $\delta_k$ clipped to $[-0.5,\, 0.5]$ to prevent extrapolation
beyond the fitting interval. Using this, we obtain the range estimate as 
\begin{equation}
\label{eq:r_para}
    \hat{r}_k^p = \left(\hat{\ell}_k + \delta_k\right) c\,T_s.
\end{equation}
This reduces the quantization error, achieving sub-sample accuracy that is well below $cT_s/2$. 

\subsection{Near-Field 2D-MUSIC}
\label{subsec:music}

The sample channel covariance matrix can be written as
\begin{equation}
\label{eq:Rk}
  \hat{\mathbf{R}}_k = \frac{1}{T}\sum \nolimits_{t=1}^{T}
    \hat{\mathbf{h}}_k^{(t)}\bigl(\hat{\mathbf{h}}_k^{(t)}\bigr)^H.
\end{equation}
From \eqref{eq:h_k_hat} and independence of $\tilde{\mathbf{w}}^t$ across $t$, we can determine that the rank of $\hat{\mathbf{R}}_k$ is ${\rm R}_k=\min(M,T)$. 
Let the  eigenvalue decomposition of channel covariance matrix be $\mathbf{R}_k=\mathbf{U}^{k}\Delta^{k}{\mathbf{U}^{k}}^H$ such that
$\mathbf{U}^{k}=[\mathbf{u}_1^{k},\dots,\mathbf{u}_{\rm R_k}^{k}]$. Clearly, the signal subspace spans the eigenvector $\mathbf{u}^{k}_1$ corresponding to dominant eigenvalue of $\mathbf{R}_k$  and the noise subspace spans the remaining eigenvectors corresponding to the smallest ${\rm R}_k-1$ eigenvalues of $\mathbf{R}_k$ denoted using $\{\mathbf{u}^{k}_m\}_{m=1}^{{\rm R}_k}$.
The 2D near-field MUSIC spectrum \cite{Wang_subspace}
is evaluated as
\begin{equation}
\label{eq:music_spec}
  P_k(r,\theta)
  = \left(\mathbf{a}_{\rm L}(r,\theta)^H
    \tilde{\mathbf{U}}^{k}{\tilde{\mathbf{U}}^{k^H}}
    \mathbf{a}_{\rm L}(r,\theta)\right)^{-1}
\end{equation}
where $\tilde{\mathbf{U}}=[\mathbf{u}_2^{k},\dots,\mathbf{u}_{\rm R_k}^{k}]$ and $\mathbf{a}_{\rm L}(r,\theta)$ is the
near-field steering vector defined in
Section~\ref{sec:system_model}.
Since the signal subspace is orthogonal to the noise subspace, $P_k(r,\theta)$ theoretically becomes unbounded at $(r,\theta)=(r_k,\theta_k)$. 
However, the accuracy of detection is subjected to the resolution of search space
$\mathcal{S}=[R_{\min}, R_{\max}]\times[0,\pi]$  used for evaluating the  MUSIC spectrum. 
Utilizing the parabolic estimate $\hat{r}_k^p$
from~\eqref{eq:r_para}, the MUSIC search can be confined to the local
range window
\begin{equation}
\label{eq:local_range_window}
  \mathcal{R}_k =
  \bigl[\hat{r}_k^p - 0.5\Delta_r,\;\hat{r}_k^p + 0.5\Delta_r\bigr],
\end{equation}
where $\Delta_r$ is the local range window. 
This allows a denser grid to be placed over a narrow window,
improving resolution at no additional cost. Thus, the location  of  $k$-th UE  can be estimated using MUSIC spectrum as
\begin{equation}
\label{eq:music_est}
  \bigl(\hat{r}_k^m,\, \hat{\theta}_k^m\bigr)
  = \operatorname*{arg\,max}_{(r,\theta)\in
    \tilde{\mathcal{S}}}~ P_k(r,\theta), \quad k = 1,\dots,K,
\end{equation}
where $\tilde{\mathcal{S}}=\mathcal{R}_k \times [0,\pi]$.
Because the search is confined to the narrow local window, the
range estimate $\hat{r}_k^m$
depends on the accuracy of
the parabolic estimate.
Thus, the accuracy of estimate $(\hat{r}_k^m,\hat{\theta}_k^m)$ remains limited by the grid spacing. Increasing the grid resolution  increases the accuracy at the cost of computational complexity.  Additionally, the accurate evaluation of sample covariance matrix  (thus, the MUSIC spectrum) depends on the extraction of LoS component. Presence of strong NLoS components or a small movement of UE may lead to inaccurate estimate of the LoS tap, which in turn may affect the accuracy of the MUSIC algorithm. 
Further, the parabolic estimate of the range parameter is limited due to inherent bias that comes by  fitting a three-point symmetric curve to an asymmetric tap energy profile.
These systematic errors motivate the deep-learning based refinement.

\subsection{Joint Refinement via NFMR-Net}
\label{subsec:mlp}
In this section, we present NFMR-Net to predict the range and angle corrections.  In particular, we  train a lightweight multi-layer perceptron~\cite{rumelhart1986learning} to predict  the range  correction $\Delta r_k$ over parabolic estimate and angle correction $\Delta\theta_k$ over MUSIC estimate. The parabolic estimate is used for correcting the range estimate since it performs better compared to the MUSIC approach, as we will see in Section \ref{sec:results}.

The range and angle carry fundamentally different physical information: range is encoded in the \emph{delay} domain through the tap energy profile, while angle is encoded in the \emph{spatial} domain through the array steering vector response.
Mixing these in shared layers for joint refinement of $\hat{r}_k$ and $\hat{\theta}_k$
requires huge representative training data set, higher dimensional model, weighted loss function that may required hyper parameter tuning to balance the trade-off between range and angle estimates. Thus, NFMR-Net
uses two separate sub-networks: the \textit{range sub-network} and the \textit{angle sub-network}. 
\subsubsection{Input representation}\label{subsub:input_representation}
Each UE feeds the following set of features into  the two sub-networks. 

 $\bullet$~\textit{Range features} $\mathbf{f}_{\rm r}$:
 ($L'+1$)-tap window of the tap energy profile centred on the
LoS tap $\hat{\ell}_k$ extracted using parabolic interpolation (see \eqref{eq:los_tap}), i.e. $\mathcal{E}_k=
\bigl[e_k[\hat{\ell}_k-1],\dots,e_k[\hat{\ell}_k+{L'}-1]\bigr]$, captures the power delay profile observed by the UE and thus allows the network to correct the  bias in parabolic estimate.
This feature vector is further supplemented by the following 
\begin{itemize}
    \item[-] parabolic range
    estimate $\hat{r}_k^p$  given in \eqref{eq:r_para},
    \item[-]  MUSIC range estimate $\hat{r}_k^m$ given in~\eqref{eq:music_est}, 
    \item[-] tap asymmetry $\tau_{\rm asym} = (e_k[\hat{\ell}_k{+}1] -e_k[\hat{\ell}_k{-}1])/e_k[\hat{\ell}_k]$ to capture NLOS-induced bias in the tap peak,
    \item[-] relative LOS dominance $\tau_{\rm peak} = e_k[\hat{\ell}_k]/\max_\ell e_k[\ell]$ to reflects multipath severity,
    \item[-] mean tap energy $\bar{e}_k=\tfrac{1}{L}\sum_{l=0}^{L-1} e_k[l]$, and 
    \item[-] quadratic phase curvature of ${\mathbf{u}}_1^k$ 
\end{itemize}
$\bullet$~\textit{Angle features} $\mathbf{f}_{\rm \theta}$:
Recall that $\mathbf{u}_1^k$ represents the dominant eigenvector of the channel covariance matrix $\hat{\mathbf{R}}_k$ given in \eqref{eq:Rk}. It captures the 1D signal sub-space that is characterized by the LoS steering vector $\mathbf{a}_{\rm L}(r_k,\theta_k)$ which inherently depends on both angle and range in the near-field. Thus, $\mathbf{u}_1^k$ exhibits the dependency on $\mathbf{a}_{\rm L}(r_k,\theta_k)$. To capture this, we define a correlation feature $ \mathbf{c}_\theta$ such that
\begin{equation}
  \mathbf{c}_\theta[i]
  = \bigl|{\mathbf{a}_{\rm L}(\hat{r}_k,\,\hat{\theta}_k^m+\vartheta_i)^H\mathbf{u}_1^{k}}
    \bigr|^2,
  \label{eq:corr_ang_mlp}
\end{equation}
where $\vartheta_i \in [-0.5\Delta_w,\,{+}0.5\Delta_w]$,  $\hat{\theta}^m_k$ is the coarse estimate of the angle determined by the MUSIC method and $\hat{r}_k$ is the  estimate of the range determined by the range sub-network. 
This fine scan is centred on $\hat{\theta}_k^m$ with local angle window $[-0.5\Delta_w,\,{+}0.5\Delta_w]$ wide enough to ensure the true angle lies within the scanned region. This will allow the angle sub-network to observe a well-defined correlation peak whose sub-grid location captures $\theta_k -\hat{\theta}_k^m$, providing fine-grained  information of angle much better as compared to grid spacing used for evaluating MUSIC spectrum.
This correlation feature vector is further supplemented by the following
\begin{itemize}
\item[-] normalised MUSIC angle estimate $\hat{\theta}_k^m$, 
\item[-] linear phase slope of $\hat{\mathbf{u}}_1^k$ , \item[-] peak correlation value $c_\theta^{\max}=\max_i |\mathbf{c}_\theta[i]|$, 
\item[-] negative normalised second derivative of $\mathbf{c}_\theta$ to capture the local fall-off/rise behavior of correlation vector,
\item[-] normalised MUSIC range estimate $\hat{r}_k^m$. 
\end{itemize}

\subsubsection{Architecture}
NFMR-Net consists of two sub-networks
applied to each of the $K$ sources independently.
Each sub-network consists of three fully-connected hidden layers (256$\rightarrow$128$\rightarrow$64)  with LayerNorm~\cite{ba2016layernormalization}, GELU
activation~\cite{lee2023geluactivationfunctiondeep}, and dropout after each hidden layer. This is followed by a
Tanh output that bounds the predicted range/angle correction to $(-1,\,1)$.
It is assumed that the sub-networks are trained offline and the weights are shared across all UEs for their simultaneous localization.  
 
\subsubsection{Loss Function and Training}
The network is trained to minimize Wing Loss~\cite{feng2018wingloss}
defined  as
\begin{equation*}
    \mathcal{L}_{\text{wing}}\left(\hat{y},y^*\right) =
    \begin{cases}
        w \ln\!\left(1 + |\hat{y} - y^*|/\varepsilon\right), & |\hat{y} - y^*| < w, \\
        |\hat{y} - y^*| - C, & \text{otherwise,}
    \end{cases}
\end{equation*}
where $\hat{y}$ and $y^*$ are the prediction and ground truth, respectively, and $C = w - w\ln(1+w/\varepsilon)$ ensures continuity.
Wing Loss is preferred over other loss functions because its logarithmic
region near zero produces larger gradients for small deviations, leading to faster convergence on fine-grained corrections.
The Wing loss is applied independently to train the 
range and angle sub-networks as
\begin{align*}
  \mathcal{L}_{\rm range}  
    &= \mathcal{L}_{\rm wing}\!\left(\widehat{\Delta r}_k,\; 
       \Delta r_k^{*}\right), \text{~and}\\
  \mathcal{L}_{\rm angle}  
    &= \mathcal{L}_{\rm wing}\!\left(\widehat{\Delta\theta}_k,\; 
       \Delta\theta_k^{*}\right),
\end{align*}
where $\Delta r_k^{*} = r_k - \hat{r}_k^p$ and 
$\Delta\theta_k^{*} = \theta_k - \hat{\theta}_k^m$ 
are the ground-truth residuals.
 
\subsubsection{Training Data}
\begin{figure}[t]
    \centering
    \includegraphics[width=0.415\textwidth]{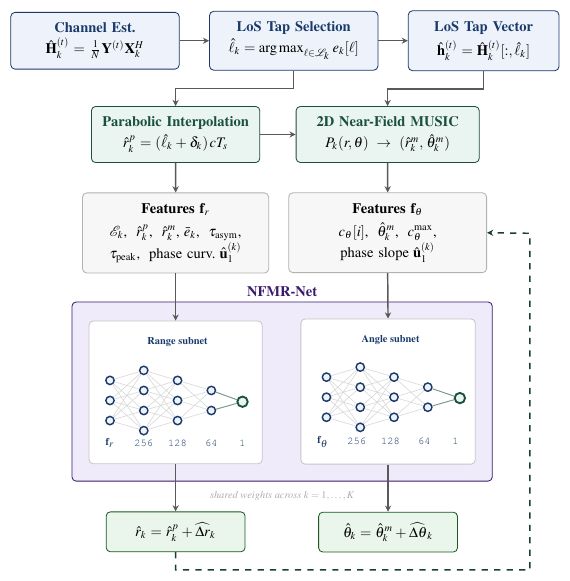}
    \caption{Proposed NFMR-Net pipeline.}
    \label{fig:pipeline}
    \vspace{-2mm}
\end{figure}

The anchor node is placed at the origin with static and uniformly distributed $P$ scatterer points. We  draw $K$  UE positions uniformly at random within the near-field region to capture the received signals for $J$ times.  
For such setup, the received signal is recorded with independent and random sampling across the parameters (like SNR, UE location, noise, etc.) to ensure wide diversity in the training dataset. Next, from the received signal, we extract the LoS component, which later is used to obtain the parabolic estimate of the range $\hat{r}_k^p$ and  MUSIC estimate of the angle $\hat{\theta}_k^m$. Finally, using these estimates, we prepare the range feature vector $\mathbf{f}_{\rm r}$ and angle feature vector $\mathbf{f}_{\rm \theta}$, as discussed in Subsection \ref{subsub:input_representation}, for each of the UE locations that are drawn in $J$ number of independent trials. 
The final estimates are recovered as
\begin{align}
  \hat{r}_k^{\rm nfmr}
    &= \hat{r}_k^p
       + \widehat{\Delta r} \text{~~and~~} \hat{\theta}_k^{\rm nfmr}
    = \hat{\theta}_k^m
       + \widehat{\Delta\theta}.
\end{align}
The complete proposed framework is illustrated in Fig. \ref{fig:pipeline}.

\section{Numerical Results and Discussion}
\label{sec:results}
In this section, we present the numerical performance analysis of proposed method and compare it with the 2D MUSIC method. 
In particular, we present the  RMSE performance of  the proposed NFMR-Net and compared it with  parabolic interpolation (Para-Int) (only for range parameter) and 2D MUSIC with and without  confined search (CS) to local range window (see \eqref{eq:local_range_window}). It may be noted that all these methods utilizes LoS vector response extracted using the  approach   proposed in Subsection \ref{subsec:channel_estimation}.
For numerical analysis, we consider the simulation parameters as given in Table \ref{table:simulation_parameters}. NFMR-Net is trained on 8000 samples and validated on 1200, and tested over a separate  set of 4000 samples. The feature vectors in each sample is recorded for $K$ simultaneous UEs. 

\begin{table}[t]
\centering
\caption{Simulation Parameters}
\label{tab:params}
\begin{tabular}{lcc}
\hline
\textbf{Parameter} & \textbf{Symbol} & \textbf{Value} \\
\hline
Carrier frequency        & $f_c$            & $2.4$\,GHz \\
Sample time              & $T_s$            & $5\times10^{-9}$\,s \\
Delay resolution         & $cT_s$           & $1.5$\,m \\
Antenna spacing          & $d_A$         & $0.5\lambda$ \\
ULA size                 & $M$              & $41$ \\
Fraunhofer distance      & $d_F$            & $100$\,m \\
Number of UEs            & $K$              & $3$ \\
Number of obstacles      & $P$              & $5$ \\
Absorption loss   & $\beta$          & $0.5$ \\
Pilot sequence length    & $N$              & $256$ \\
Number of transmissions      & $T$              & $4$ \\
Local range window       & $\Delta_r$       & $8$\,m \\
Local angle window          & $\Delta_w$        & $10^\circ$ \\
SNR range                & —                & $[-5,\,20]$\,dB \\
UE range                 & —                & $[5,\,30]$\,m \\
UE angle                 & —                & $[0^\circ,\,180^\circ]$ \\
\hline
\end{tabular}\label{table:simulation_parameters}
\end{table}

\begin{figure}[t!]
    \subfloat{\includegraphics[width=0.23\textwidth]{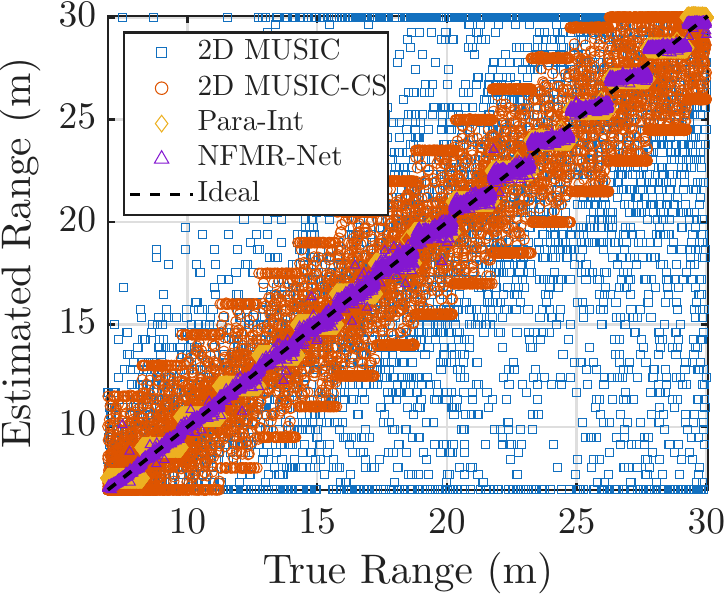}}
    \hfill
    \subfloat{\includegraphics[width=0.23\textwidth]{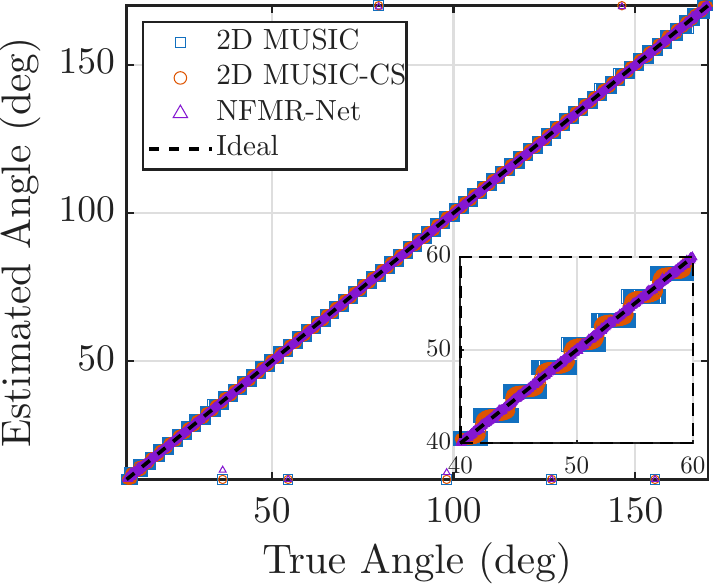}}
    \vspace{-2.5mm}
    \caption{Scatter plots of range (left) and angle (right) parameters.}
    \label{fig:scatter}
    \vspace{-2mm}
\end{figure}
Fig. \ref{fig:scatter} provides the scatter plot of the range and angle estimates as a function of their true values for SNR range $0$ to $20$ dB. Fig. \ref{fig:scatter}(left) shows that the error in the range estimates of both MUSIC algorithms are much higher as compared to those of the proposed parabolic or NFMR-Net methods. Further, the Fig.~\ref{fig:scatter}(right) shows that the angle estimate errors of 2D MUSIC is much lower and confined to the used angular grid resolution. However, it can be seen that the proposed NFMR-Net provide much better angular refinement over the estimates of MUSIC methods.

\begin{figure}[t!]
    \subfloat{\includegraphics[width=0.23\textwidth]{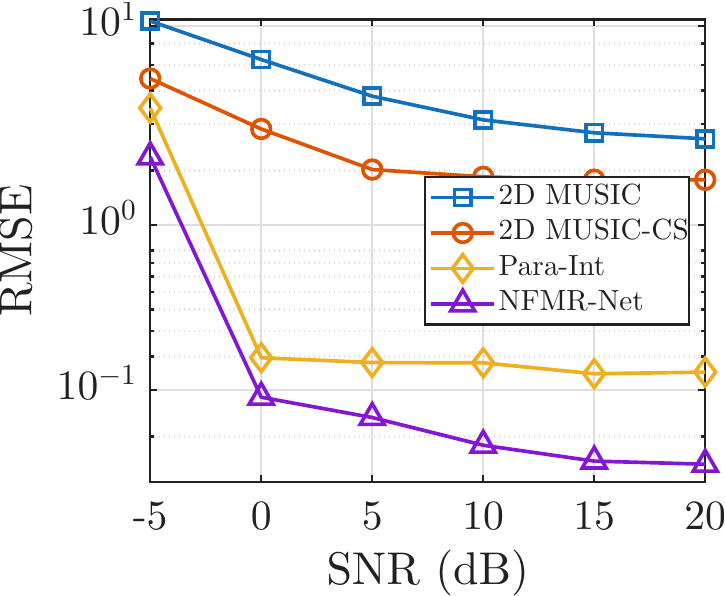}}
    \hfill
    \subfloat{\includegraphics[width=0.23\textwidth]{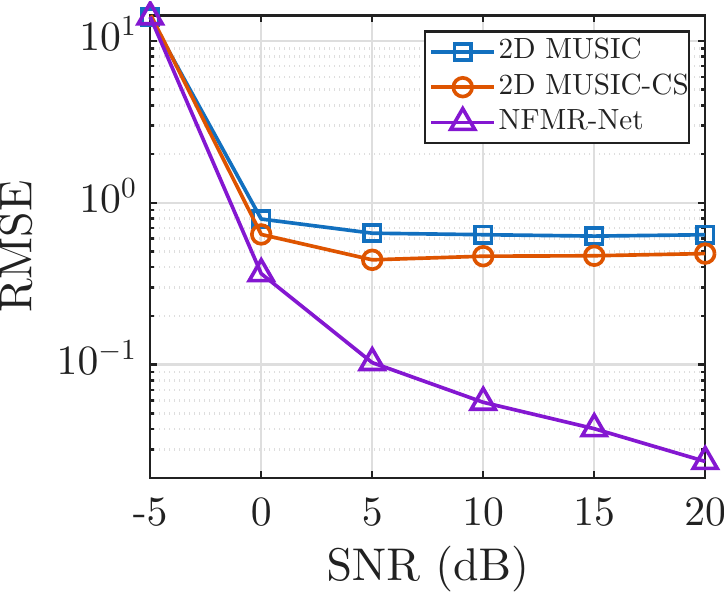}}
    \vspace{-2.5mm}
    \caption{RMSE vs. SNR for range (left) and angle (right).}
    \label{fig:snr}
    \vspace{-2mm}
\end{figure}

Fig.~\ref{fig:snr} shows that the RMSE performances of the range and angle parameters improve with the increase in SNR, as expected. Fig.~\ref{fig:snr}(left) shows that the parabolic interpolator
outperforms both MUSIC variants. This  confirms that delay-domain energy profile based interpolation is more suitable to extract the range information in near-field as compared to the  subspace based  methods. Besides,  the poor range estimation performance of MUSIC methods  is due to the fact that the spherical wavefront in the near-field makes channel response to be sensitive to the  range. The RMSE of the parabolic interpolation saturates at high SNR because of the limited delay resolution. Nevertheless, the proposed NFMR-Net achieves the lowest  RMSE across wide range of SNR, as it learns to correct the residual bias for given static scenario that parabolic interpolation could not remove. Further, Fig.~\ref{fig:snr}(right) shows that the performance of both MUSIC methods converge to an error floor with increase in SNR. This occurs because of  the static and finite angular grid spacing, implying that the error can not reduce below the resolution limit. 
NFMR-Net, on the other hand, continues to improve the RMSE of angle estimate with the increase in SNR, leveraging the  correlation profile in neighborhood of coarse estimate (see ~\eqref{eq:corr_ang_mlp}) to extract fine-grain angular information that could not be captured by MUSIC because of finite grid resolution. 

\begin{figure}[t!]
    \subfloat{\includegraphics[width=0.23\textwidth]{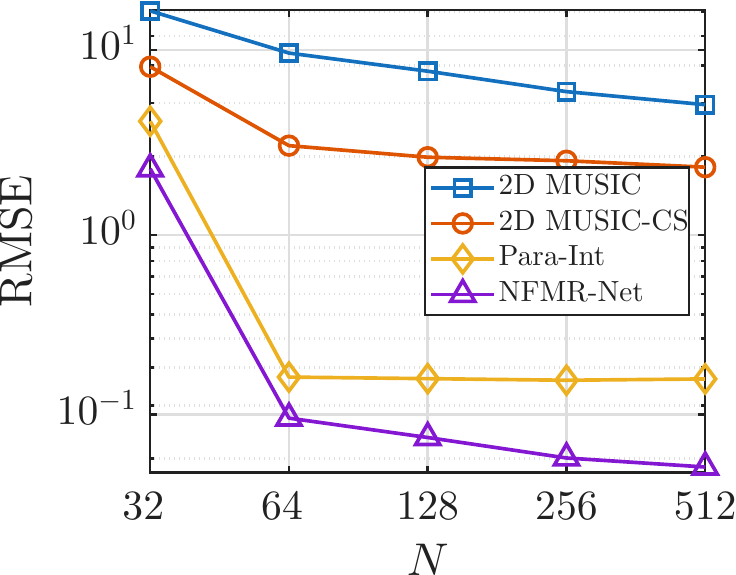}}
    \hfill
    \subfloat{\includegraphics[width=0.23\textwidth]{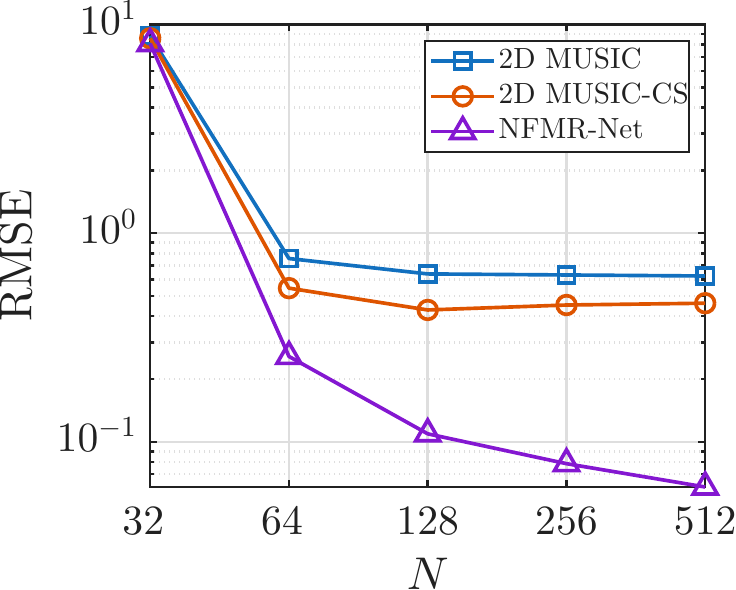}}
    \vspace{-2.5mm}
    \caption{RMSE vs. $N$, for range (left) and angle (right), at  SNR $=10$ dB.}
    \label{fig:rmse_vs_n}
    \vspace{-2mm}
\end{figure}

Fig. \ref{fig:rmse_vs_n} shows that the RMSE  of all methods improves with the increase in the length of pilot $N$. This is quite evident because the inter-user interference (see \eqref{eq:Hk_hat}) decreases with the increase in $N$. This is due to ZC pilot sequences with distinct roots  assigned to UEs become almost orthogonal as $N$ increases. While the MUSIC methods improve initially and saturate, the proposed NFMR-Net  achieves the lowest RMSE and improves consistently with the increase in $N$.

\section{Conclusion}
This paper presents a deep learning framework, called NFMR-Net, for  wideband near-field multi user localization. The proposed framework first extracts the LoS array response and  corresponding delay-tap for each UE from the multi-tap channel matrix by leveraging the properties of CP added ZC pilot sequences. Using this, we next obtain the coarse estimates of range and  angle using parabolic interpolation and MUSIC, respectively. The residual errors in these coarse estimates are then minimized using NFMR-Net, which employs two separate MLPs for range and angle.
Numerical results demonstrate that the NFMR-Net achieves significantly lower  RMSE as compared to the conventional 2D-MUSIC. 
The RMSE of NFMR-Net  decreases monotonically with the increase in SNR and  ZC sequence length. However, 2D MUSIC exhibits saturation  because of the
quantization error resulting from 
finite grid spacing.   
\bibliographystyle{IEEEtran}  

\end{document}